Title:      K-/K+ ratios in relativistic heavy-ion collisions
    Authors:    G.Q. Li and G.E. Brown

    Dear Editor,

    We would like to submit the above paper to Phys. Rev. C for
   possible publication in Section: Relativistic Heavy-Ion
   Collisions. Enclosed please find tone LaTeX file and 17 postscript
   files for figures. Please send further correspondence to

        G.Q. Li
        Dept. of Phys. and Astron.
        SUNY-Stony Brook, NY 11794
        phone: 516-632-8136
        fax:   516-632-9718
        email: gqli@nuclear.physics.sunysb.edu

    with best regards,

     Guoqiang

\documentstyle[aps,epsfig,twocolumn]{revtex}

\begin{document}

\draft 


\title{$K^-/K^+$ ratios in relativistic heavy-ion collisions}

\bigskip
\author{G.Q. Li and G.E. Brown}
\address{Department of Physics and Astronomy, State University of New
York at Stony Brook,\\
Stony Brook, New York 11794}
\maketitle
  
\begin{abstract}
We study $K^-/K^+$ ratios as a function of centrality
(participant nucleon number), transverse mass ($m_t$), 
and rapidity, in heavy-ion collisions at beam energies
between 1A and 2A GeV. We use the relativistic transport model that
includes expicitly the strangeness degrees of freedom
and consider two scenarios for kaon properties in
dense matter, one with and one without medium modifications
of their properties. In both scenarios, The $K^-/K^+$ ratio 
does not change very much with the centrality, while the 
$K/\pi$ and ${\bar K}/\pi$ ratios increase with increasing centrality. 
Significant differences are predicted, both in magnitudes and 
shapes, for the $m_t$ spectra and rapidity distributions
of $K^-/K^+$ ratio. Experimental measurement of these ratios, 
currently under investigation by the FOPI, KaoS, E866, and E895 
collaborations, will be useful in revealing the kaon in-medium properties. 

\end{abstract}


\narrowtext

\section{Introduction}

Whether and how hadronic properties, such as their masses,
widths, and dispersion relations, are modified in hot and 
dense medium is a topic of great current interest. 
Of particular importance are the medium modifications of kaon 
properties, as they are related to both spontaneous and explicit
chiral symmetry breaking, and they are useful inputs for the
study of kaon condensation and neutron star properties
\cite{brown88a,thor94}. Since the pioneering work of Kaplan 
and Nelson \cite{kap86} on the possibility of kaon condensation 
in nuclear matter, a huge amount of theoretical effort has been 
devoted to the study of kaon properties in dense matter,
using such diversified approaches as chiral perturbation
theory \cite{brown87,wise91,brown94,lee95,kai95,waas96,lee96,kai97,waas97}, 
the Nambu$-$Jona-Lasinio model \cite{lutz94}, and SU(3) Walecka-type 
mean-field model \cite{sch94,knor95}. Although quantitatively 
results from these different models are not identical, qualitatively, 
a consistent picture has emerged; namely, in nuclear matter
the $K^+$ feels a weak repulsive potential, whereas
the $K^-$ feels a strong attractive potential. 

Experimentally, in-medium properties of kaon and antikaon
can be obtained from the analysis of kaon-nucleus scattering
\cite{brown88,fried97} and kaonic atom data \cite{gal94}. 
The information so obtained is, unfortunately, restricted 
to low densities. For the study of kaon condensation, densities 
much higher than that accessible by kaonic atoms are involved.
This can only be obtained by analysing heavy-ion collision 
data on various observables involving kaon and antikaon. 
Measurements of kaon spectra and flow have been systematically
carried out in heavy-ion collisions at SIS (1-2 AGeV), 
AGS (10 AGeV), and SPS (200 AGeV) energies \cite{qm96}. 
The analysis of their yields, spectra, and in particular collective
flow has indeed provided useful information about kaon
properties in dense nuclear matter 
\cite{likoli95,liko95a,liko96,cass97,brat97,fae97a,fae97b,lilee97a,lilee97b}.
So far most of the experimental data from the FOPI \cite{ritman,fopi,best97}
and KaoS \cite{kaos94,elm96,kaos96,kaos97} collaborations at SIS/GSI 
seem to be consistent with the predictions from the chiral perturbation 
theory.

To put these conclusions on a firmer footing, additional observables
and experimental data from independent collaborations will
certainly be useful. In this paper we study the centrality, 
transverse mass ($m_t$), and rapidity dependence of the $K^-/K^+$ 
ratio in heavy-ion collisions. Since the medium effects act 
oppositely on kaon and antikaon, their ratio can reflect more 
precisely these effects, as was pointed out in 
Refs. \cite{ogil97,ogil98}. We will consider Ni+Ni, Ru+Ru and
Au+Au collisions at beam energies between 1 and 2A GeV. 
These systems are being investigated 
by the FOPI \cite{herr} and KaoS \cite{peter} collaborations 
at SIS/GSI. Furthermore, the Au+Au collisions are also being 
analysed by the E866 \cite{ogil97,ogil98} and E895 \cite{eos,e895}
collaborations at the AGS/BNL.

This paper is arranged as follows. In Section II, we briefly
review the relativistic transport model, kaon in-medium
properties and elementary kaon production cross sections. 
The results are presented in Section III. The paper ends 
with a short summary in Section IV.

\section{The relativistic transport model and kaon production}

Heavy-ion collisions involve very complicated nonequilibrium
dynamics. One needs to use transport models in order to
extract from experimental data the information about 
in-medium properties of hadrons. In this work we will
use the relativistic transport model similar to that 
developed in Ref. \cite{ko87}. Instead of the usual linear 
and non-linear $\sigma$-$\omega$ models, we base our model on 
the effective chiral Lagrangian recently developed by Furstahl, 
Tang, and Serot \cite{fst}, which is derived using dimensional 
analysis, naturalness arguments, and provides a very good description 
of nuclear matter and finite nuclei. In the mean-field approximation, 
the energy density for the general case of asymmetric nuclear matter 
is given by
\begin{eqnarray}
\varepsilon _N& = & {2\over (2\pi )^3} \int _0^{K_{fp}} 
d{\bf k} \sqrt {{\bf k}^2+m_N^{*2}}\nonumber\\
 & + &{2\over (2\pi )^3} 
\int _0^{K_{fn}} d{\bf k} \sqrt {{\bf k}^2+m_N^{*2}} \nonumber\\
 & + & W\rho +R {1\over 2}(\rho_p-\rho_n) -{1\over 2C_V^2}W^2
- {1\over 2C_\rho^2}R^2 + {1\over 2C_S^2}\Phi^2 \nonumber \\
 & +& {S^{\prime 2}\over 4C_S^2}d^2\left\{\left(1-{\Phi \over S^\prime}
\right)^{4/d}\left[{1\over d}{\rm ln}\left(1-{\Phi \over S^\prime}
\right) - {1\over 4}\right]+{1\over 4}\right\} \nonumber\\
 & -& {\xi\over 24}W^4 - {\eta \over 2C_V^2}{\Phi \over S^\prime}W^2. 
\end{eqnarray}
The nucleon effective mass $m_N^*$ is related to its scalar
field $\Phi$ by $m_N^*=m_N-\Phi$. $W$ and $R$ 
are the isospin-even and isospin-odd vector potentials,
respectively. The last three terms give the self-interactions of
the scalar field, the vector field, and the coupling between
them. The meaning and values of various parameters in Eq. (1)
can be found in \cite{fst}.

From the energy density of Eq. (1), we can also derive a 
relativistic transport model for heavy-ion collisions. 
At SIS energies, the colliding system consists mainly of 
nucleons, delta resonances, and pions. While medium effects 
on pions are neglected, nucleons and delta resonances propagate 
in a common mean-field potential according to the Hamilton 
equation of motion,
\begin{eqnarray}
{d{\bf x}\over dt} = {{\bf p}^*\over E^*}, \;\;\;
{d{\bf p}\over dt} = - \nabla _x (E^*+W),
\end{eqnarray}
where $E^*=\sqrt {{\bf p}^{*2} + m^{*2}}$.
These particles also undergo stochastic two-body
collisions, including both elastic and inelastic scattering. 

In heavy-ion collisions at incident energies considered in
this work, kaons can be produced from pion-baryon and 
baryon-baryon collisions. For the former we use cross sections 
obtained in the resonance model by Tsushima {\it et al.} 
\cite{fae94}. For the latter the cross sections obtained in 
the one-boson-exchange model of Ref. \cite{liko95b,likoc98} 
are used. Both models describe the available experimental data 
very well. For antikaon production from pion-baryon collisions 
we use the parameterization proposed by Sibirtsev {\it et al.} 
\cite{sib97}. For baryon-baryon collisions, we use a somewhat different 
parameterization, which describes the experimental data better,
than Ref. \cite{sib97}. In addition, the antikaon
can also be produced from strangeness-exchange processes
such as $\pi Y\rightarrow {\bar K}N$ where $Y$ is either a
$\Lambda$ or $\Sigma$ hyperon. The cross sections for
these processes are obtained from the reverse ones, ${\bar K}N\rightarrow 
\pi Y$, by the detailed-balance relation. All the parameterizations
for the elementary cross sections and comparisons with experimental data    
can be found in our recent paper \cite{lilee97b}.

Particles produced in elementary hadron-hadron 
interactions in heavy-ion collisions cannot 
escape the environment freely. Instead, they
are subjected to strong final-state interactions.
For the kaon, because of strangeness conservation,
its scattering with nucleons at low energies is
dominated by elastic and pion production processes,
which do not affect its final yield but changes its momentum 
spectra. The final-state interaction for the antikaon is much 
stronger. Antikaons can be destroyed in the strangeness-exchange 
processes. They also undergo elastic scattering. 
Both the elastic and absorption cross sections increase
rapidly with decreasing antikaon momenta. This will have
strong effects on the final $K^-$ momentum spectra in heavy-ion
collisions.

we will consider two scenarios for kaon properties in nuclear
medium, one with and one without medium modification.  
From the chiral Lagrangian the kaon and antikaon in-medium energies
can be written as \cite{lilee97b}
\begin{eqnarray}
\omega _K=\left[m_K^2+{\bf k}^2-a_K\rho_S
+(b_K \rho )^2\right]^{1/2} + b_K \rho 
\end{eqnarray}
\begin{eqnarray}
\omega _{\bar K}=\left[m_K^2+{\bf k}^2-a_{\bar K}\rho_S
+(b_K \rho )^2\right]^{1/2} - b_K \rho 
\end{eqnarray}
where $b_K=3/(8f_\pi^2)\approx 0.333$ GeVfm$^3$, 
$a_K$ and $a_{\bar K}$  are two parameters
that determine the strength of the attractive scalar potential for
kaon and antikaon, respectively. If one considers only the
Kaplan-Nelson term, then $a_K=a_{\bar K}=\Sigma _{KN}/f_\pi ^2$.
In the same order, there is also the range term which acts differently
on kaon and antikaon, and leads to different scalar attractions.
Since the exact value of $\Sigma _{KN}$ and the size of
the higher-order corrections are still under intensive
debate, we take the point of view that $a_{K,{\bar K}}$ can be treated 
as free parameters and try to constrain them from the experimental 
observables in heavy-ion collisions. In Ref. \cite{lilee97a,lilee97b} 
we found that $a_K\approx 0.22$ GeV$^2$fm$^3$ and $a_{\bar K}\approx 
0.45$ GeV$^2$fm$^3$ provide a good description of kaon and antikaon
spectra in Ni+Ni collisions at 1A and 1.8A GeV. These values will
be used in this work as well.

\begin{figure}
\begin{center}
\centerline{\epsfig{file=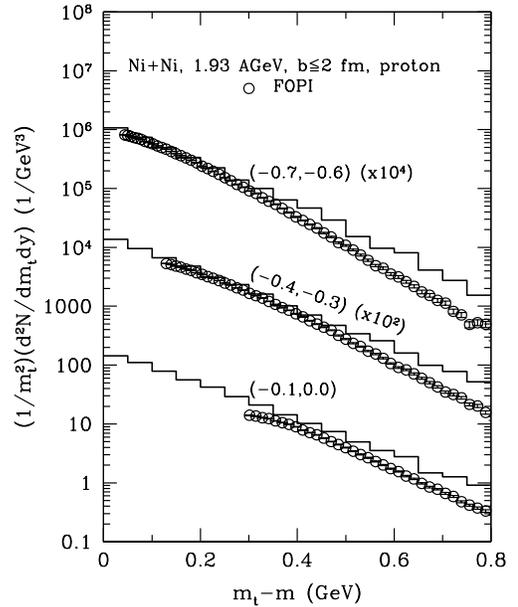,height=3.5in,width=3.5in}}
\caption{Proton transverse mass spectra in central
Ni+Ni collisions at 1.93 AGeV in three rapidity bins. The open
circles are experimental data from the FOPI collaboration
\protect\cite{hong97,hong98}.}
\end{center}
\end{figure}

In nuclear medium, kaon and antikaon masses are modified,
so are their production thresholds. We will then use
$\sqrt {s_0^*}$, which are calculated with effective
masses, in evaluating the in-medium production cross sections.
This amounts to the change of threshold, or approximately,
to the change of available phase space. 
In addition to the change in the production cross sections,
the medium effects on kaon and  antikaon also affect their 
momentum spectra, when they propagate in the mean-field potentials.
The Hamilton equations of motion for kaon and antikaon are very
similar to those for nucleons \cite{liko95a},
\begin{eqnarray}
{d{\bf r}\over dt} = {{\bf k}\over \omega _{K,{\bar K}} \mp b_k\rho _N},
\;\; {d{\bf k}\over dt} = - \nabla _x U_{K,{\bar K}},
\end{eqnarray}
where the minus sign corresponds to kaon, and the plus sign to antikaon.
It is clearly that the $K^+$ momentum increases and that of $K^-$
decreases when they propagate in their respective mean field potentials.
This affects significantly their momentum spectra, and especially
the momentum spectra of their ratio. 

\begin{figure}[hbt]
\begin{center}
\centerline{\epsfig{file=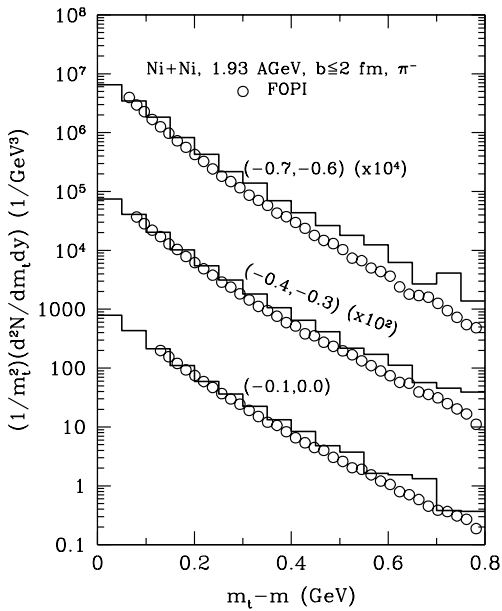,height=3.5in,width=3.5in}}
\caption{Same as Fig. 1, for $\pi^-$.}
\end{center}
\end{figure}

\section{Results and discussions}

In this section we present our results for the centrality, 
transverse mass, and rapidity dependences of $K^-/K^+$ ratio. 
Before that, we first compare the proton and pion $m_t$ spectra 
from our calculation with the available experimental data, 
taking central Ni+Ni collisions at 1.93A GeV as an example.
In Fig. 1 we compare the proton transverse mass spectra in 
three rapidity bins with the experimental data from the FOPI 
collaboration \cite{hong97,hong98}. Similar comparison for
the $\pi^-$ transverse mass spectra is shown in Fig. 2.
Our results are seen to be in good agreement with the
data \cite{hong97,hong98}. Similarly, in Ref. \cite{lilee97b}
the proton and $\pi^-$ rapidity distributions obtained in our 
transport model were shown to be in good agreement with the FOPI data.

\subsection{centrality dependence}

In this subsection, we discuss the centrality dependence of
the $K^-/K^+$ ratio. We use the participant nucleon number
$A_{part}$ as the measurement of the centrality. 
In Fig. 3 we show the $A_{part}$ dependence of the
pion multiplicity in Ni+Ni collisions at 1.93A GeV.
The solid line gives 1/3 of the total pion number obtained
in our calculation, while the circles and squares are
the FOPI data \cite{pelte97} for $\pi^-$ and $\pi^+$,
respectively. It is seen that our results are in good agreement with
the data, and that the pion multiplicity increases almost 
linearly with the participant nucleon number. In other words,
the $N_\pi /A_{part}$ ratio is almost independent of the centrality.

\begin{figure}[hbt]
\begin{center}
\centerline{\epsfig{file=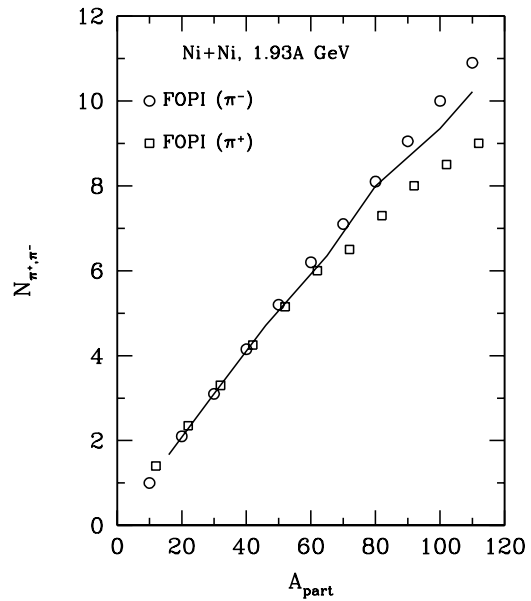,height=3.5in,width=3.5in}}
\caption{Pion multiplcity as a function of the participant
nucleon number in Ni+Ni collisions at 1.93 AGeV.}
\end{center}
\end{figure}

On the other hand, it is well-known that the $K^+$ 
(usually shown in terms of $K^+/\pi^+$ ratio) yield increases
more than linearly with the $A_{part}$ for beam energies 
ranging from 1A GeV \cite{kaos94,elm96,kaos96,likoc98} to 10A GeV 
\cite{ogil97,ahle96}. Our results for $K/\pi$ and ${\bar K}/\pi$ 
ratios in Au+Au collisions are shown in Fig. 4. It is seen that either
with or without kaon medium effects, these ratios increase
more than linearly with the participant nucleon number.
This is due to the increasing importance of the secondary
processes involving baryon-resonances, pions, and hyperons in the
case of antikaon production, when going from peripheral to central
collisions. The E866 data for Au+Au collisions at 11.6A GeV/c indicate 
that $K^-/\pi^+$ ratio also increases with increasing 
centrality \cite{ahle96}. 
 
\begin{figure}[hbt]
\begin{center}
\centerline{\epsfig{file=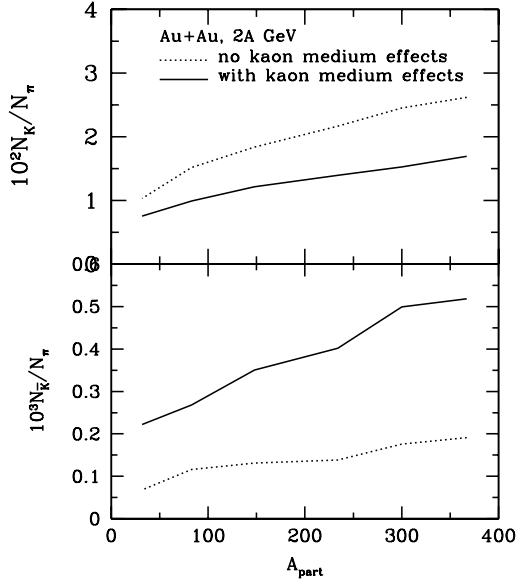,height=3.5in,width=3.5in}}
\caption{$K/\pi$ and ${\bar K}/\pi$ ratios as a function of the 
participant nucleon number in Au+Au collisions at 2A GeV.}
\end{center}
\end{figure}

The centrality dependence of the $K^-/K^+$ ratios is shown in
Figs. 5, 6, and 7, for Ni+Ni at 1.8A GeV, Ru+Ru at 1.69A GeV,
and Au+Au at 2A GeV,
respectively. For Ni+Ni collisions, we also shown in Fig. 5
the experimental data from the KaoS collaboration \cite{kaos97}
by open circles. For all the systems considered,
it is seen that either with or without medium effects,
the $K^-/K^+$ ratio does not depend very much on the centrality,
since in both scenarios, the $K/\pi$ and ${\bar K}/\pi$
ratios increase at about the same rate with increasing
centrality (see Fig. 4). The experimental data of the KaoS
collaboration for Ni+Ni collisions (Fig. 5) show little
increase of the ratio towards central collisions within
their statistical uncertainties. Preliminary data from the
FOPI collaboration for Ru+Ru at 1.69A GeV \cite{leif97}, and data
from E866 collaboration for Au+Au collisions at 10.6A GeV
\cite{ogil97} both show weak centrality dependence for 
the $K^-/K^+$ ratio.
 
\begin{figure}[hbt]
\begin{center}
\centerline{\epsfig{file=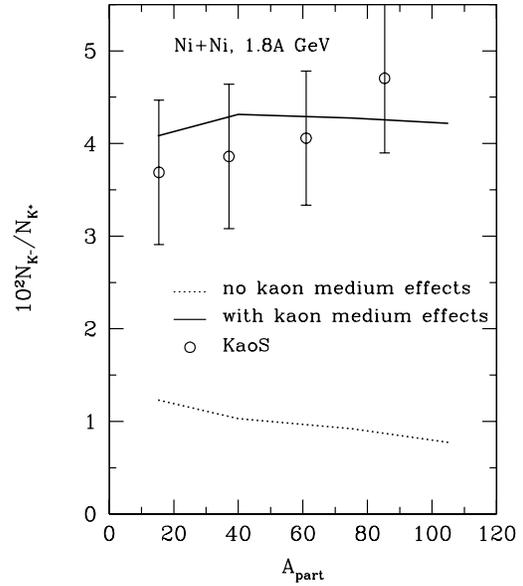,height=3.5in,width=3.5in}}
\caption{$K^-/K^+$ ratio as a function of the 
participant nucleon number in Ni+Ni collisions at 1.8A GeV.
The open circles are the experimental data from the
KaoS collaboration \protect\cite{kaos97}}
\end{center}
\end{figure}

\begin{figure}
\begin{center}
\centerline{\epsfig{file=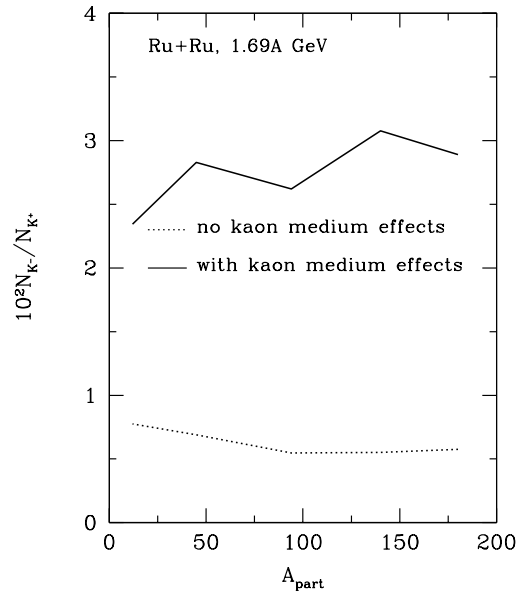,height=3.5in,width=3.5in}}
\caption{Same as Fig. 5, for Ru+Ru collisions at 1.69A GeV.}
\end{center}
\end{figure}

Without kaon medium effects, the $K^-/K^+$ ratio ranges from 
0.006 in Ru+Ru at 1.69A GeV, to 0.01 in Ni+Ni collisions at 1.8A GeV, 
to 0.0075 in Au+Au collisions at 2A GeV. The increase of this ratio
from Ru+Ru to Ni+Ni comes from the increase in beam energy, 
since at these energies the antikaon excitation function
is steeper than that of kaon because of a higher threshold.
On the other, the decrease of this ratio from Ni+Ni to Au+Au
is due to the fact that in the large Au+Au system the antikaon
absorption effects become more significant. Qualitatively, the
absorption probability is proportional to the product
of the absorption cross section and the average size of the
system. In this case the experimental data from the KaoS collaboration
for Ni+Ni collisions are significantly underestimated (see Fig. 5).

\begin{figure}[hbt]
\begin{center}
\centerline{\epsfig{file=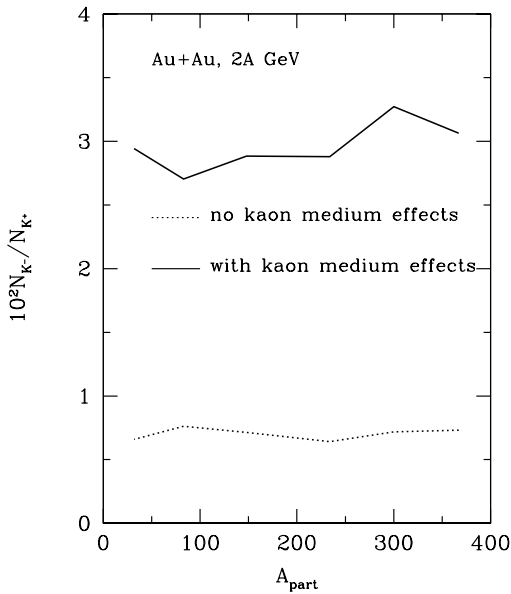,height=3.5in,width=3.5in}}
\caption{Same as Fig. 5, for Au+Au collisions at 2A GeV.}
\end{center}
\end{figure}

When kaon medium effects are included, the $K^-/K^+$ ratio
increase by about a factor of 4, for all the systems
considered here. This increase results from a factor of 3
increase in antikaon yield and about 35\% reduction in
kaon yield. In this case, the $K^-/K^+$ ratio from the KaoS
collaboration can be nicely explained. Furthermore,
preliminary data from the FOPI collaboration
indicate a $K^-/K^+$ ratio of about 0.02-0.03 in Ru+Ru
collisions at 1.69A GeV. Our predictions that include kaon medium 
effects are seen to be in better agreement with these preliminary
data than those without the kaon medium effects. 
 
Naively, it is expected that when kaon medium effects are included,
the $K^-/K^+$ ratio should increase with the increasing centrality,
since in central collisions reduction of the antikaon mass and the
increase of kaon mass are the most significant. However, since
the second processes $Y\pi\rightarrow {\bar K}N$ play an important
role in antikaon production, and hyperon yield, which are produced
in association with kaons, is reduced most significantly in central 
collisions, the increase in the reduction of antikaon mass towards
central collisions is largely compensated. The increase
of the $K^-/K^+$ ratio with increasing centrality when kaon 
medium effects are included is thus marginal (on the order of 10-20\%).

It is also of interest to show the beam energy dependence
of the $K^-/K^+$ ratio in central Au+Au collisions. This is
done in Fig. 8 for impact parameter b= 1 fm. In both
scenarios with and without kaon medium effects, the
ratio is seen to increase as beam energy increases.
This is understandable, as the antikaon production
threshould is higher than that of kaon, so at these
energies, the antikaon production cross sections 
increases faster than that of kaon.

Finally, we show in Fig. 9 the time evolution of central
density, kaon yield, and antikaon yield in Au+Au collisions
at 1.5A GeV and b=3 fm. It is seen that for a considerable 
duration of time the system is compressed to a density of
around 3$\rho_0$, and during this period of time, most of
the kaons and antikaons are produced. As was shown in
Ref. \cite{lilee97b}, $K^-$ condensation is predicted to
occur around three times normal nuclear matter based
on current theoretical and empirical information. The
experimental data on kaon and antikaon yields, spectra,
ratios, and flow from Au+Au collisions at 1.5A Gev
will  be very useful in pinning down the
critical density for kaon condensation.

\begin{figure}[hbt]
\begin{center}
\centerline{\epsfig{file=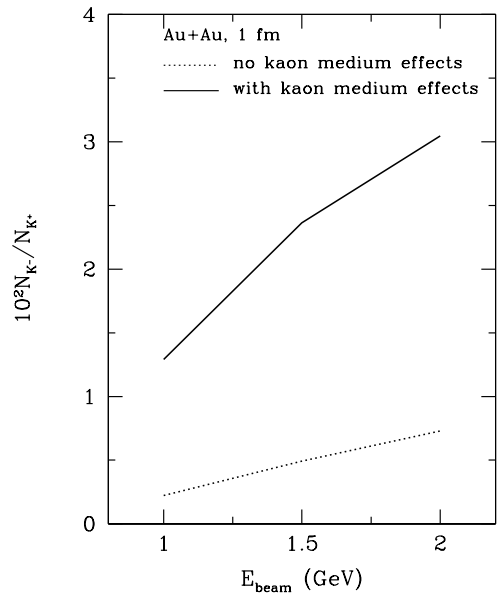,height=3.5in,width=3.5in}}
\caption{Beam energy dependence of $K^-/K^+$ ratio
in central Au+Au collisions.}
\end{center}
\end{figure}

\begin{figure}[hbt]
\begin{center}
\centerline{\epsfig{file=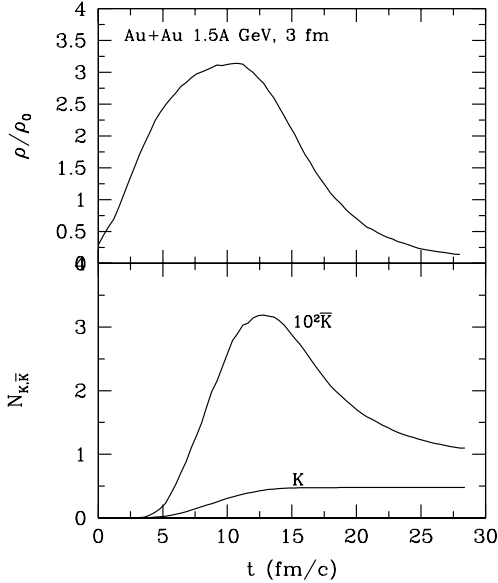,height=3.5in,width=3.5in}}
\caption{Time evolution of central density as well as kaon and antikaon
yields in Au+Au collisions at 1.5A GeV and 3 fm.}
\end{center}
\end{figure}

\subsection{Transverse mass spectra}

The effects of kaon and antikaon mean-field potentials 
can be more clearly seen by looking at their ratio as a 
function of their tranverse mass (or kinetic energy). The results are shown
in Figs. 10, 11, 12, and 13, for Ni+Ni at 1.8A GeV,
Ru+Ru at 1.69A GeV, Au+Au at 1.5A GeV,  and Au+Au at 2A GeV,
respectively. The results for Ni+Ni are presented
in terms of kinetic energy for minimum-biased collisions,
in accordance with the experimental data from the
KaoS collaboration \cite{kaos97}, shown in Fig. 10 with open
circles. 

When kaon medium effects are neglected, the $K^-/K^+$ ratio
is seen to increase slightly with increasing transverse 
momentum. Since the antikaon absorption cross section by nucleons
becomes large at low momentum, low-momentum antikaons are more 
strongly absorbed than high-momentum ones. This makes the $K^-/K^+$ 
ratio decrease at small transverse mass. When medium effects are
included, we find that the shape of the $K^-/K^+$ ratio 
is completely different from that without kaon medium effects.
The ratio is now seen to increase dramatically towards small
transverse mass. For example, for Ru+RU collisions, it 
increases from about 0.02 at $m_t-m_K=0.3$ GeV
to about 0.08 when $m_t-m_K$ approaches zero. 
The difference in the shape of the $m_t$ spectra comes
from the propagation of kaons and antikaons in their
mean-field potential. Kaons are `pushed' to high momenta by 
the repulsive potential, while antikaons are `pulled' to 
low momenta, leading to an enhanced $K^-/K^+$ ratio
at small transverse masses. From Fig. 10, we also
see that the effects of propagation in mean-field potentials
are more pronounced in central collisions.

\begin{figure}[hbt]
\begin{center}
\centerline{\epsfig{file=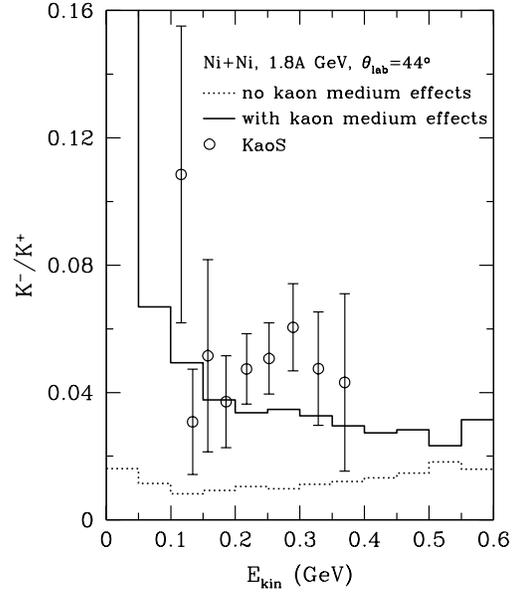,height=3.5in,width=3.5in}}
\caption{Kinetic energy spectra of the $K^-/K^+$ ratio in 
Ni+Ni collisions at 1.8A GeV. The experimental data are from the
KaoS collaboration \protect\cite{kaos97}}
\end{center}
\end{figure}

\begin{figure}[hbt]
\begin{center}
\centerline{\epsfig{file=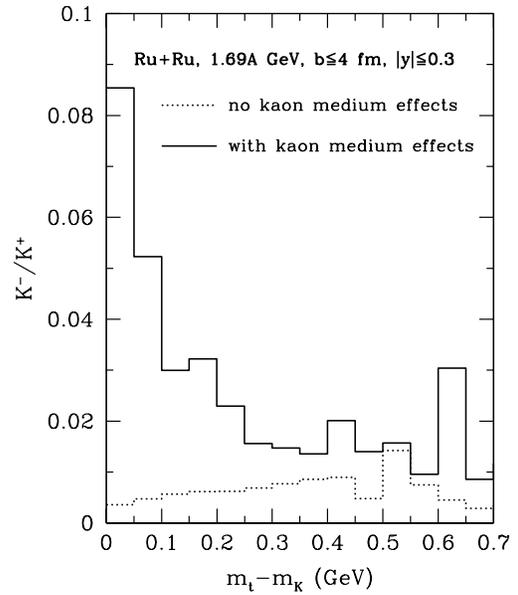,height=3.5in,width=3.5in}}
\caption{Transverse mass spectra of $K^-/K^+$ ratio in central
Ru+Ru collisions at 1.69A GeV.}
\end{center}
\end{figure}

\begin{figure}[hbt]
\begin{center}
\centerline{\epsfig{file=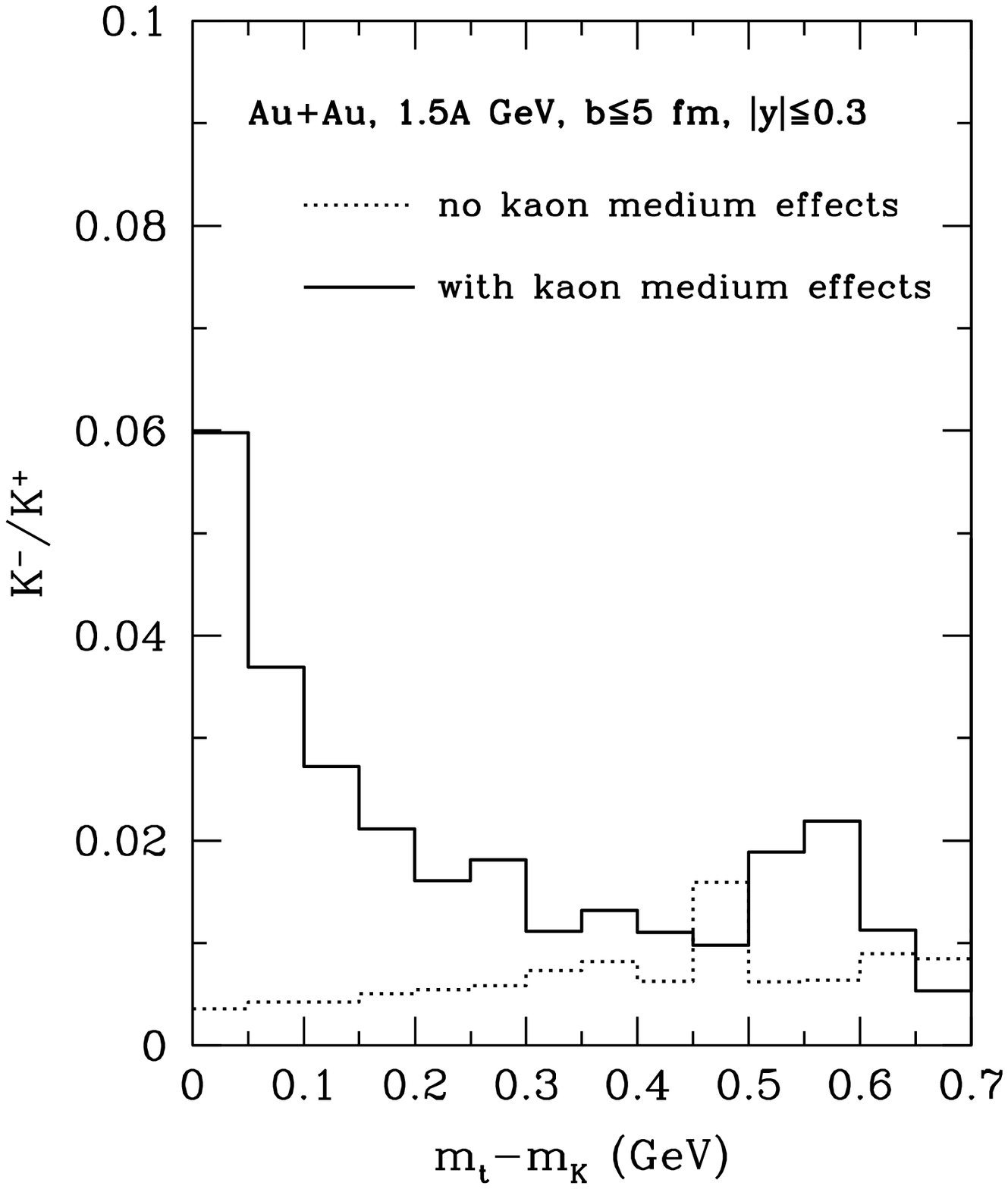,height=3.5in,width=3.5in}}
\caption{Same as Fig. 11, for Au+Au collisions at 1.5A GeV.}
\end{center}
\end{figure}

\begin{figure}[hbt]
\begin{center}
\centerline{\epsfig{file=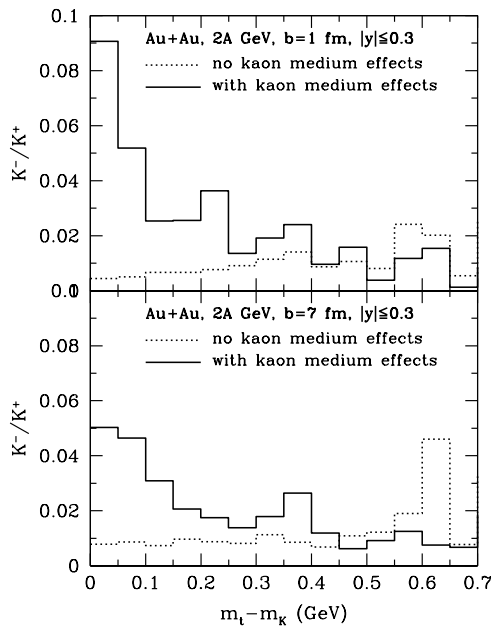,height=3.5in,width=3.5in}}
\caption{Same as Fig. 11, for Au+Au collisions at 2A GeV.}
\end{center}
\end{figure}

Experimental data from the KaoS collaboration provided
some indication for this dramatic increase of the
$K^-/K^+$ ratio at small momenta \cite{kaos97}. Our results
are compared with these data in Fig. 10. 
It is seen that our results including the
kaon medium effects are in much better agreement with
the data. Of course the statistical uncertainty of the
data is still quite large (especially for the first datum point
at 0.12 GeV). This will be improved in their recent analysis
of Ni+Ni collisions at 1.93A GeV \cite{peter}. It will also be 
very useful if the ratio at kientic energies less than 0.1 GeV 
can be measured experimentally.

It should be emphasized that in this work, as in our previous study of
kaon production in heavy-ion collisions, the explicit momentum
dependence of the kaon scalar and vector potential is not considered.
This should be a reasonable approximation for heavy-ion collisions
at 1-2A GeV, as kaons (antikaon) produced in these reactions usually
have small momenta. This approximation will break down at higher
beam energies about 10A GeV. A recent theoretical calculation 
\cite{lutz97} indicated that the attractive antikaon potential 
becomes weaker as its momentum (relative to the medium) increases. 
Experimental data on $K^-$-nucleus scattering at high incident
momenta also indicate a repulsive (rather than an attractive) antikaon
optical-model potential. This momentum dependence of the kaon
(antikaon) potential might explain the fact that so far the
experimental data for heavy-ion collisions at 10A GeV do not show substantial
enhancement of the $K^-/K^+$ ratio approaching small transverse
masses (momenta).

\subsection{Rapidity distribution}

The rapidity distributions of the $K^-/K^+$ ratio should provide
quite similar information on kaon medium effects as
its transverse mass spectra. The results for Ni+Ni at 1.93A Gev,
Ru+Ru at 1.69A GeV, Au+Au at 1.5A GeV,
and Au+Au collisions are shown in Figs. 14, 15, 16, and 17, respectively.
For Ni+Ni collisions we show also the preliminary data from
the FOPI collaboration \cite{leif97,hong98b} by circles.

Indeed, the magnitudes and the shapes of these rapidity 
distributions with and without kaon medium effects are
very different. Without kaon medium effects, the
rapidity distribution is seen to decrease slightly from target/projectile
rapidities to mid-rapidity. This is again due to the
large absorption cross section for slow-moving antikaons.
When kaon medium effects are
included, there appears not only an overall increase of the
ratio, because of the increased production cross section of
antikaons, but also a significant change of the shape
of the rapidity distribution. The $K^-/K^+$ ratio is
seen to increase steadily from target/projectile rapidities
to mid-rapidity. Our results with kaon medium effects
are in much better agreement with the preliminary data
from the FOPI collaboration (Fig. 14).

\begin{figure}[hbt]
\begin{center}
\centerline{\epsfig{file=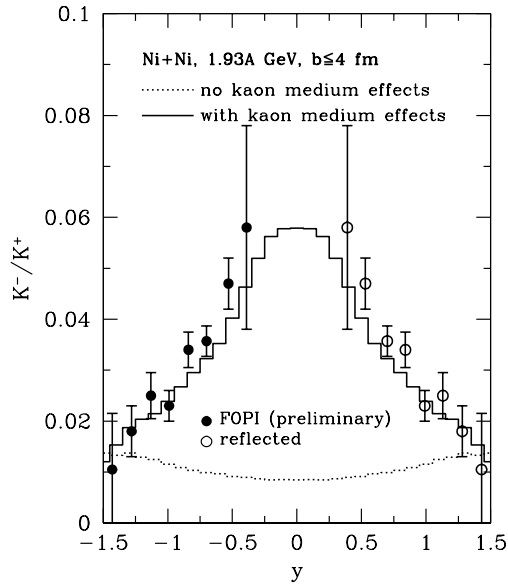,height=3.5in,width=3.5in}}
\caption{Rapidity distribution of $K^-/K^+$ ratio in central
Ni+Ni collisions at 1.93A GeV. The solid circles are 
preliminary experimental data from the FOPI collaboration 
\protect\cite{leif97,hong98b}, while the open circles are
obtained by reflecting the data with respect to the mid-rapidity.}
\end{center}
\end{figure}

\begin{figure}[hbt]
\begin{center}
\centerline{\epsfig{file=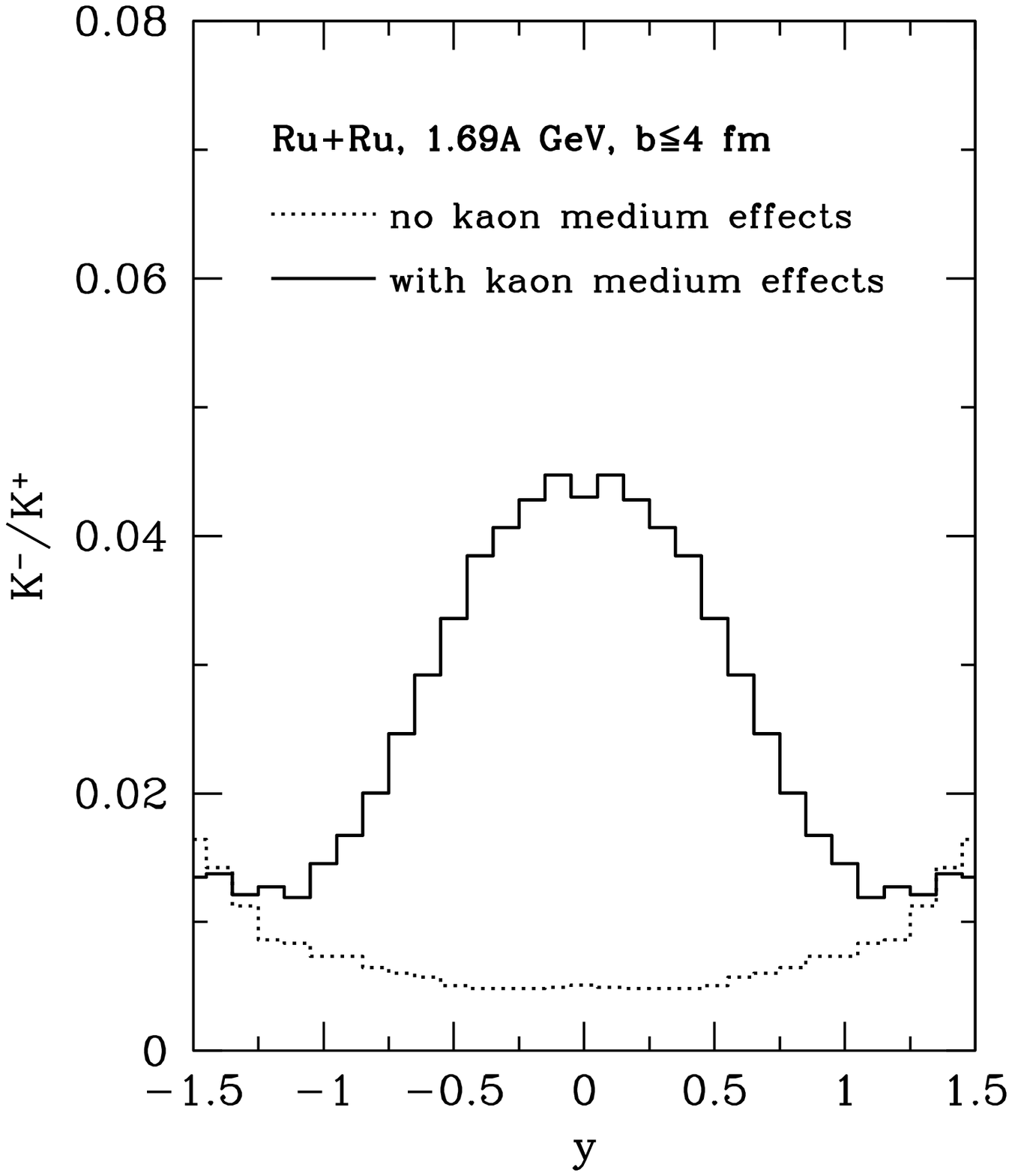,height=3.5in,width=3.5in}}
\caption{Same as Fig. 14, for Ru+Ru collisions at 1.69A GeV.}
\end{center}
\end{figure}

\begin{figure}[hbt]
\begin{center}
\centerline{\epsfig{file=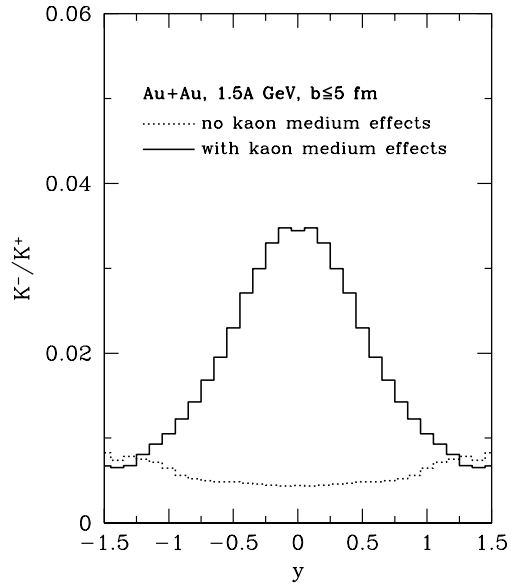,height=3.5in,width=3.5in}}
\caption{Same as Fig. 14, for Au+Au collisions at 1.5A GeV.}
\end{center}
\end{figure}

For Au+Au collisions at 2A GeV and $b=1$ fm, we
also show in the figure the $K^-/K^+$ ratio before
any final-state interactions on kaons and antikaons
are included, which means that even the antikaon 
absorption is turned off. We see that in this case, the
ratio also decreases from mid-rapidity to projectile
and target rapidities. The decrease of this ratio towards
large rapidities (momenta) reflects the fact that
antikaons are produced at a higher threshold than kaons,
so that their momentum distributions are restricted
more severely by the available energies. However,
we need to emphasize that antikaon absorption and the
fact that its absorption cross section increases at low
momenta are well-known experimental facts and must be
included in transport model calculations.  

As mentioned in the Introduction, so far most of the
experimental data from the FOPI and KaoS collaborations
at SIS/GSI are consistent with the chiral perturbation theory
predictions for kaon in-medium properties. For an ultimate confirmation
of these medium effects, it should be be very useful if
independent experimental data for Au+Au collisions at 2A GeV 
from the E866 and E895 collaborations could become available.

\begin{figure}[hbt]
\begin{center}
\centerline{\epsfig{file=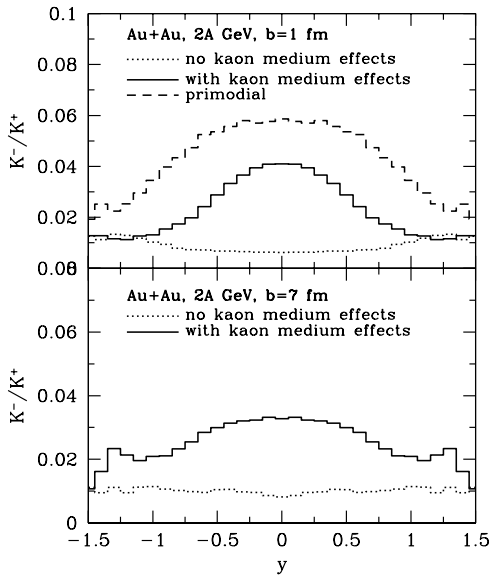,height=3.5in,width=3.5in}}
\caption{Same as Fig. 14, for Au+Au collisions at 2A GeV.}
\end{center}
\end{figure}

\section{Summary}

In summary, we studied $K^-/K^+$ ratios as a function of centrality
(participant nucleon number), transverse mass ($m_t$), 
and rapidity, in Ni+Ni, Ru+Ru, and Au+Au collisions at beam 
energies between 1 and 2A GeV. We used the relativistic transport 
model that includes explicitly the strangeness degrees of freedom
and considered two scenarios for kaon properties in
dense matter, one with and one without medium modifications
of their properties. In both scenarios, The $K^-/K^+$ ratio 
does not change very much with the centrality, while the 
$K/\pi$ and ${\bar K}/\pi$ ratios increase with increasing centrality. 
Significant differences were predicted, both in magnitudes and 
shapes, for the $m_t$ spectra and rapidity distributions
of $K^-/K^+$ ratio. We found that the experimental data from
the KaoS collaboration for the kinetic energy spectra of
the $K^-/K^+$ ratio and those from the FOPI collaboration
for its rapidity distribution support the suggestion of kaon
medium effects. We emphasize that the independent data from the 
E866 and E895 collaborations for Au+Au collisions at 2A GeV
will be useful in confirming or confronting these findings. 

\vskip 0.5cm

We are grateful to N. Herrmann and P. Senger for sending us
the data files, and to N. Herrmann, C. Ogilvie, and P. Senger for
useful communications. This work is supported in part by the 
Department of Energy under Grant No. DE-FG02-88ER40388.

\end{document}